# A Query-based Routing Table Update Mechanism for Content-Centric Network

Pei-Hsuan Tsai
Institute of Manufacturing Information
and Systems
National Cheng Kung University
Tainan, Taiwan
phtsai@mail.ncku.edu.tw

Yu-Lin Tseng
Institute of Manufacturing Information
and Systems
National Cheng Kung University
Tainan, Taiwan
p96054147@ncku.edu.tw

Jun-Bin Zhang
Department of Computer Science and
Information Engineering
National Cheng Kung University
Tainan, Taiwan
p78083025@gs.ncku.edu.tw

Meng-Hsun Tsai
Department of Computer Science and
Information Engineering
National Cheng Kung University
Tainan, Taiwan
tsaimh@csie.ncku.edu.tw

*Abstract*—Due to the popularity of network applications, such as multimedia, online shopping, Internet of Things (IoT), and 5G, the contents cached in the routers are frequently replaced in Content-Centric Networking (CCN). Generally, cache miss causes numerous propagated packets to get the required content that deteriorates network congestion and delay the response time of consumers. Many caching strategies and routing policies were proposed to solve the problem. This paper presents an alternative solution by designing a query-based routing table update mechanism to increase the accuracy of routing tables. By adding an additional query content in interest packets, our approach real-time explores the cached content in routers and updated the routing table accordingly. This paper uses a general network simulator, ndnSIM, to compare basic CCN and our approach. The results show that our approach improves the response time of consumers and network congestion and is compatible with general forwarding strategies.

*Keywords—content-centric networking, routing table, query name, exploration, update mechanism*

## I. Introduction

In recent years, with the continuous development and change of network applications, the amount of network data has grown rapidly, and the network system is more complex and difficult to maintain. The current TCP/IP (Transmission Control Protocol/Internet Protocol) network architecture, faces insufficient IP address to support. Therefore, several new network architectures have been proposed, such as Named Data Networking (NDN) [1], Content-Centric Networking (CCN) [2-3].

CCN is an innovative Information-Centric Networking (ICN) derivative architecture [4-7]. Traditional IP networks tell users where to get data from IP addresses, while CCN uses content names to replace IP addresses. It focuses on what data the user wants to obtain. In CCN, the content name is expressed in the form of a hierarchical structure prefix, and the prefix of each layer is connected by "/", which is similar to the URL in the Internet [8]. Unlike TCP/IP networks, CCN allows content to be temporarily stored on the router. Therefore, most consumers can obtain the requested content directly from the routers, instead of the content producer. This greatly reduces network packet traffic and the waiting time of consumers.

In CCN, due to limited Cache Storage (CS) space and the numerous data in the network, the cached content is frequently replaced but the routing tables are not updated in real-time resulting in continuous cache miss. Cache miss brings erroneous interest packets forwarding and retransmissions that causes the network congestion and postpone the waiting time of consumers.

To reduce the cache miss in CCN, some researches focus on caching strategy and routing policies. Caching strategies based on probability [9-10] or popularity [11-12] are adopted by many traditional researches. These strategies are considered implicit because it is independently decided by the router and does not require additional packet traffic [6]. The implicit approach reduces cache redundancy and increases the cache hit rate. But it ignores non-path cached content and cached popular content may be replaced [13]. The potential of in-network caching and multi-path forwarding cannot be fully exploited [13].

Therefore, some studies have proposed explicit approaches, such as actively broadcasting to notify neighboring routers [6-14], coordinating neighboring routers to cache [15], establishing a management system to distribute data [16-17], and sending packets to explore [18]. The explicit approaches utilizing the off-path cached content [13]. However, the disadvantage of explicit approaches is the inevitable additional increase in network packet traffic and computational cost [13].

This paper presents a query-based routing table update mechanism. Interest packets and data packets are added with a field of query name, respectively. When the interest packet is forwarded, select the name of the most popular entry in the Pending Interest Table (PIT) and add it as the query name to the interest packet. The router receives the interest packet that looks up the requested content, and also look up the query name in the CS. The query result is returned through the data packet. The query-based mechanism allows routers to explore the cached content in routers. The query result was recorded in the Forwarding Information Base (FIB). The update mechanism of FIB takes into account the response time of obtaining the content and the length of its existence in the FIB. In order to timely replace the invalid routing information and ensure the accuracy of the routing information in the FIB.

The main contributions of this paper are as follows:

- Different from cache miss improvements, this paper improves network congestions by updating routing tables through exploring the contents of routers.
- Instead of continuously and actively querying routers, which increases the cost of the network. Our mechanism reduces this cost.
- Our approach is compatible with general forwarding strategies and easy to be implemented.

The organization of this paper is as follows. Section 2 introduces some concepts of CCN and the forwarding process of packets in the basic CCN. In Section 3, the improved interest packet and data packet processing process, and the routing table update mechanism are explained in detail. Section 4 introduces the parameter settings of the network simulator. Section 5 compares and analyzes the results under the two test conditions of smart flooding and best router. The conclusion part summarizes this work.

## II. OVERVIEW OF CONTENT-CENTRIC NETWORK

### A. Content-Centric Network

Interest packet and data packet used as the information exchange carrier in CCN. Consumers send out interest packets, and the interest packets forwarded to other routers or terminal servers according to the routing information in the routing table. After obtaining the requested content, a data packet returned to the consumer along the interest packet forwarding path.

As shown in Fig. 1, the interest packet contains Name, Selector, and Nonce. Name is the name of the content requested by the consumer. Selector is used to attach filtering conditions to limit the scope of interest packet forwarding, or to filter content. Nonce is a random number that is to prevent the router from receiving the same interest packet and entering the routing loop. If the interest packet has the same nonce record in the routing table, it means that the interest packet has been forwarded by this router before, and the data packet has not yet been sent back. The router will discard the repeated interest packet and no longer forward it. The data package includes Name, Signature, Signature Info, and Data. Signature is the key to the content, which means signing information. Signature Info is information related to the key, such as the creator or the timestamp of the content.

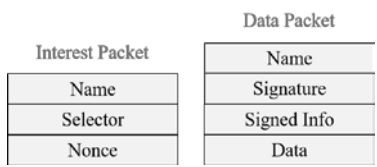

Fig. 1. CCN packet format.

There are three very important routing tables in CCN: CS, PIT, FIB, as shown in Fig. 2. The CS is the router's cache memory, which is used to cache content. So that consumers can obtain content from nearby routers in the future. The CS updates the cached content according to the cache replacement policy. Commonly used cache replacement policies include Least Recently Used (LRU), Least Frequently Used (LFU) [19], and First In First Out (FIFO).

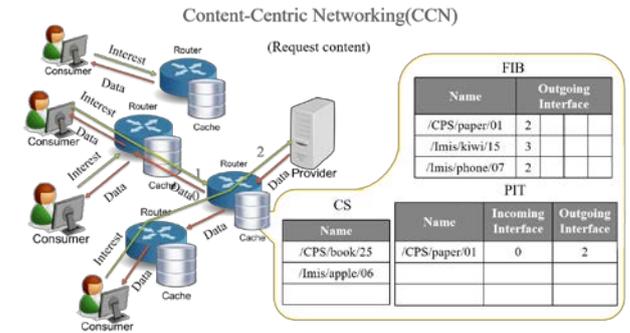

Fig. 2. Content-Centric Network architecture diagram.

The PIT is used to record the interest packet information that has been forwarded by the router but has not been satisfied and is waiting for the requested content. An entry in the PIT contains the name of the requested content, the incoming interface for receiving the interest packet, and the outgoing interface for forwarding the interest packet. If the router receives an interest packet with the same content, it will merge into the same entry. After the router receives the data content, it forwards the data packet to the consumer according to the entry information in the PIT.

The FIB is similar to the current IP routing table and is used to record the routing path information. The FIB entry includes content name and outgoing interface for forwarding of interest packets. A FIB entry can record multiple outgoing interfaces. The router will decide how to forward the interest packet according to the forwarding strategy, so as to obtain the content efficiently.

### B. Basic CCN Packet Forwarding Process

The interest packet forwarding process is shown in Fig. 3 [20]. After the router receives the interest packet, it first checks whether it has matched the requested content in the CS. If matched, the data packet directly return to the consumer. If not matched, the router continues to check whether there is an entry with the same name as the requested content in the PIT. If not, it means that the router is not waiting for the content response, and PIT adds a new entry to record the content name and the incoming interface of the interest packet. Then the router decides how to forward the interest packet or how many interest packets to forward to the next-hop router according to the forwarding strategy. The interface of the interest packet forwarding is recorded in the outgoing interface of the PIT entry. If the PIT already has an entry with the same content name as the interest packet, it means that the router has received the interest packet with the same content before and is waiting for the content response. At this time, the router only needs to record the incoming interface in the PIT and is no need to forward the interest packet to other routers.

The data packet forwarding process is shown in Fig. 4 [18, 20]. After receiving the returned data packet, the router forwards the data packet according to the incoming interface in the PIT entry and deletes the PIT entry at the same time. Then, the FIB records the routing information. The FIB first searches for entries that match the content name in the data packet. If not matched, the FIB add an entry with that name. If have matched the FIB entry, the router only needs to record the outgoing interface of the data packet. If the entry has the same outgoing interface, the router only needs to update the metric of this outgoing interface. Metric is defined as response time to obtain the requested content after forwarding the interest packet. Multiple outgoing interfaces in the FIB entry are sorted according to metric. The smaller the metric value, the higher the ranking, and the shorter the response time of this outgoing interface. The forwarding strategy preferentially forwards the interest packet to the outgoing interface with the highest ranking. Finally, the router decides whether to cache this content in the CS according to the cache replacement policy.

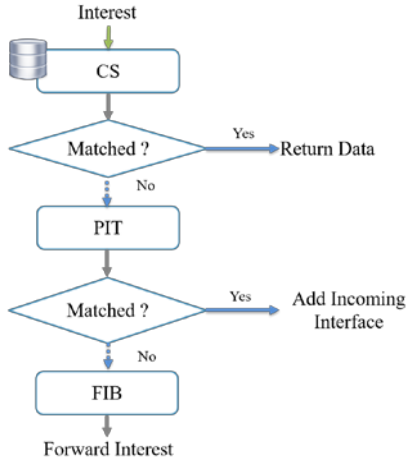

Fig. 3. CCN interest packet transfer process.

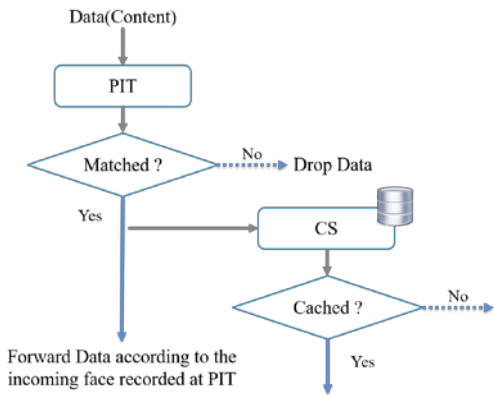

Fig. 4. CCN data packet transfer process.

### III. IMPROVEMENT MECHANISM

#### A. Improvement Packet Forwarding Process

As shown in Fig. 5, the improvement mechanism in this paper is to add a query name field to the interest packet and the data packet respectively. Use interest packets to query whether routers have other cached content. The query result is returned to the router through the data packet. Consumers send interest packets, and the first router that the interest packets arrive at will still give priority to check whether has matched the requested content in the CS. If no matched, before the interest packet is forwarded to other routers, the router selects a currently most popular content name from the PIT incoming interface and adds it to the interest packet.

The improvement interest packet forwarding process is shown in Fig. 6. After the router receives the improvement interest packet, it checks first whether has matched the requested content by the consumer in the CS. If no match, it records the interest packet information in the PIT and forwards the interest packet according to the FIB forwarding strategy. If the requested content is cached in the CS, the router check again whether there is cached the query name. If not cached the query name, the data packet containing the requested content is directly returned. If the query name is cached in the CS, and the remaining survival period is relatively long. In other words, the query content is in the first half of the caching sort in the CS, indicating that it not be replaced in the short term. When returning the data package of the requested content, the query result is added to the data package.

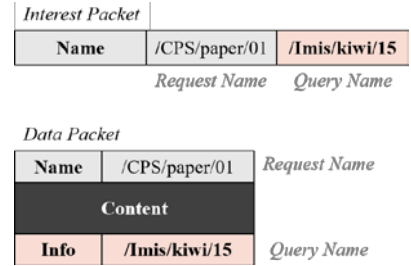

Fig. 5. Packet format of the improvement mechanism.

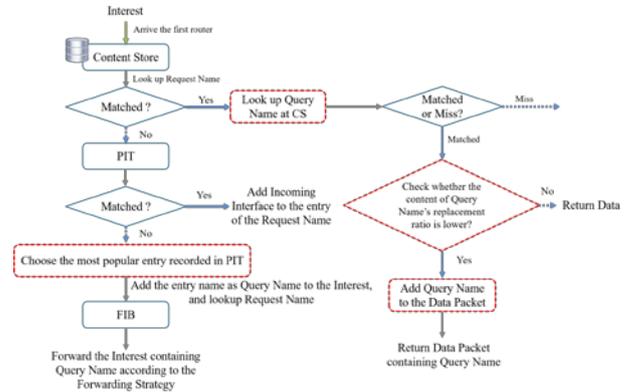

Fig. 6. Improvement interest packet forwarding process.

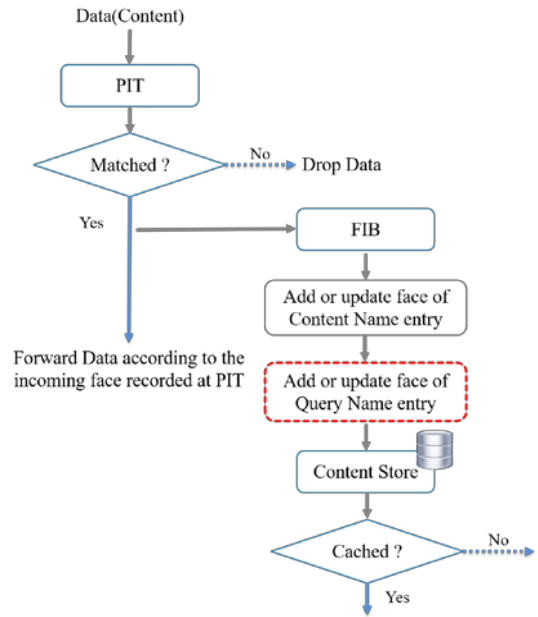

Fig. 7. Improvement data packet forwarding process.

The improvement data packet forwarding process is shown in Fig. 7. After the router receives the data packet, in addition to record the routing information of the requested content in the FIB, and also to record the routing information of the query name in the FIB.

*B. Routing Table Update Mechanism*

The routing table update mechanism presented in this paper takes the response time of the new routing information and the time of the routing information in the FIB (which refers to the difference between the time when the routing information is recorded and the current time) also taken into consideration. Fig. 8 shows the pseudo-code of the routing table update mechanism. When routers receive the data packet and need to record new routing information if the number of interfaces recorded by the same entry in the FIB is full. Compare the response time of the received data packet with the response time of the last green state interface in the FIB. The last green interface means that the waiting time to obtain content from this interface is the longest in this entry. If the response time of the newly added interface is shorter than it, replace it with the newly added interface. That is to change its state from green to yellow, and the state of the newly added interface to green. Otherwise, check all the interfaces in the FIB for routing information that has not been updated for a long time and needs to be replaced. If the time difference between the last time the interface received the data packet and the current time has exceeded the threshold, the state of that interface is changed from green to yellow. The threshold is defined as

$$T = \frac{C}{F} \quad (1)$$

where $T$ is the threshold, $C$ is the cache size, and $F$ is the amount of content generated per second. The content corresponding to the interface that exceeds the threshold $T$ has a high probability of being replaced, causing the routing information to become invalid. replaced. So the state of this interface is changed from green to yellow.

```
Algorithm. UpdateEntryFace
1. function UpdateEntryFace ( contentName, newFace, responseTime, receiveDataTime)
2.   if list-green.length ≥ maximum of faces in the list-green then
3.     if responseTime < reponseTime of last face in the list-green then
4.       replace the last face in the list-green
5.   else
6.     for each face in list-green
7.       if face of (Now - receiveDataTime) > T_n then
8.         turn status of face to yellow
9.       end if
10.    end for
11.    if list-green.length < maximum then
12.      turn the status of newFace to green
13.    end if
14.  end if
15. end if
16. end function
```

Fig. 8. Routing table update mechanism.

## IV. SIMULATION EXPERIMENT

This paper uses NS-3 based ndnSIM [21] simulation development platform to realize the network simulation environment of CCN and NDN architecture. As shown in Fig. 9, Abilene topology is used to structure the network node distribution.

Abilene topology is a high-bandwidth network with 12 routers in total, and each router represents a city. In this experiment, each router is set as a producer that can publish content belonging to its city prefix. For example, Seattle will provide data with the prefix "/Sea". At the same time, each router is also configured as a consumer that sends 100 interest packets to the network every second. In the experiment, the interest packets sent by consumers have a total number of 1,200 about the content names. The experiment simulation duration is 180 seconds.

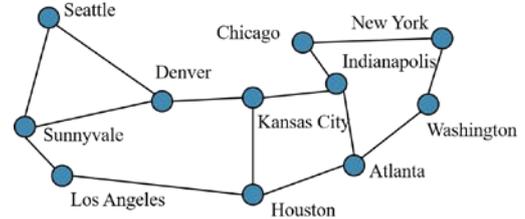

Fig. 9. Abilene topology.

In this experiment, the threshold $T$ is set to the top 50% for cached content ranking in CS. The timer time of the PIT entry is set according to the response time of the green state interface in the FIB corresponding entry. If there is no green state interface to select when forwarding the interest packet, then set the timer time of this PIT entry to the initialization value (2 seconds). According to the experimental topology, the number of green-state interfaces is set to 2. A basic CCN simulator was also established using ndnSIM as a comparative experiment. The detailed experimental parameter settings are shown in Table Ⅰ.

TABLE I. Experimental parameter settings

| Parameter | Description |
| --- | --- |
| Simulator | ndnSIM |
| Topology | Abilene topology |
| Forwarding strategy | smart flooding, best route |
| Cache replacement policy | FIFO, LFU, LRU |
| Number of routers | 12 |
| Cache size | 480-840 chunks |
| Frequency | 100 interest/seconds |
| Number of contents | 1200 |
| Survival time of query name | 50% of cache |
| Maximum of faces in list-green | 2 |
| Simulation time | 180 seconds |

## V. RESULTS AND DISCUSSIONS

The storage space of the router's CSes has a direct huge impact on the performance of the CCN [18]. In this paper, we evaluate the performance of basic CCN and our optimization mechanism, for different normalized cache sizes from 0.4 to 0.7. The results are as follows:

*A. Smart flooding*

In the simulator, smart flooding is used for our optimization mechanism and the basic CCN forwarding strategy. Under the conditions of different cache replacement policies (such as LRU, LFU, FIFO), the number of flooding, the number of interest packets, and the average response time of obtained the content are tested.

The purpose is to compare the performance impact of our approach and basic CCN due to invalid or un-updated routing information. The result is shown in Fig. 10, and the abscissa represents a different normalized cache size.

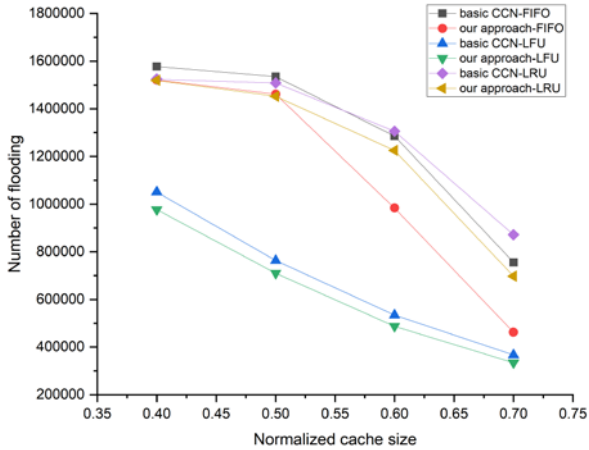

Fig. 10. The number of flooding for smart flooding.

As can be seen from Fig. 10, as the normalized cache size increases, the number of flooding for three cache replacement policies gradually decreases. The reason is that the router is able to cache more content and the cached content in CS is not easily replaced. The routing information in the FIB is also not easily invalidated. Our approach allows the FIB to record more routing information by deleting the green state interface that has not been updated for a long time. This increases the number of green status interfaces and improves the accuracy of the routing information. The result is that the number of flooding in our approach is less than basic CCN. When the cache size is small, the difference in the number of flooding between LRU and FIFO is very small. The small cache size means that the cached content has high variability and is relatively easy to be replaced, which causes the routing information in the FIB to become invalid. Fig. 11 shows the number of interest packets. Due to the small number of flooding in our approach, the number of interest packets is also relatively small.

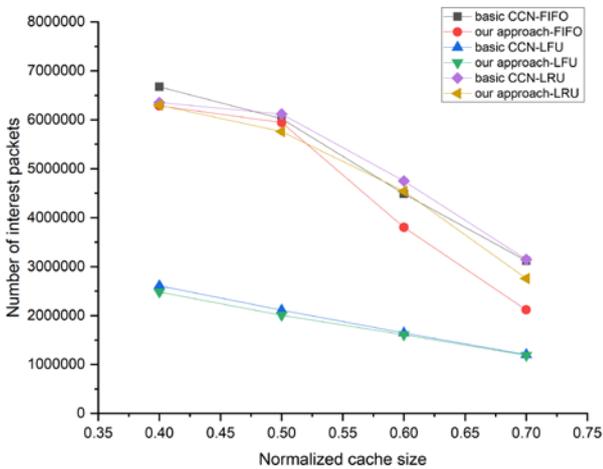

Fig. 11. The number of interest packets for smart flooding.

Fig. 12 shows the average response time for all routers in CCN to obtain content responses after forwarding interest packets. When the cache size is small, the cached content in the CS is easy to be replaced, which causes the routing information in the FIB to become invalid, cause false forwarding, and increase response time. Our approach can ensure that the routing information in the FIB is updated in time, thereby reducing the response time.

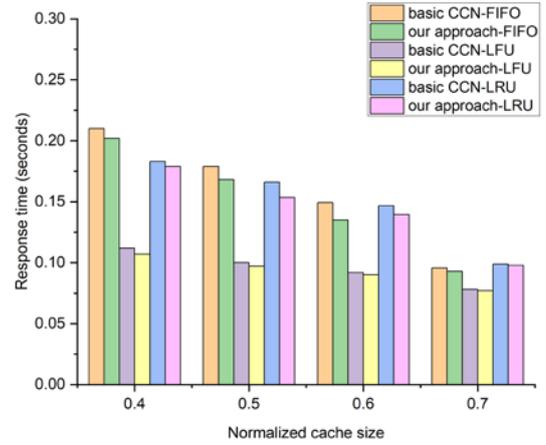

Fig. 12. Average response time for smart flooding.

### B. Best route

In the case that our approach and the basic CCN adopt the best route forwarding strategy, the experiment tested the number of retransmissions of LRU, LFU, and FIFO. It is used to measure the accuracy of the routing information in the FIB. When the PIT timer responds, the router does not receive the data packet with the content, which causes retransmission. The FIB forwards the interest packet to the next interface. The fewer the number of retransmissions, the more accurate the routing information of the green status interface is recorded in the FIB.

As shown in Fig. 13, the number of retransmissions in our approach is much smaller than basic CCN. This is because in the case of LFU, the variability of the old cached content in the CS is lower, and the new cached content is easier to be replaced. The new routing information in the FIB is more likely to become invalid and cause continuous retransmissions. But our approach has added the function of a query, which can increase the routing information in the FIB. And these inquiries routing information are within the top 50% of the CS ranking, ensuring the accuracy of the routing information in the FIB.

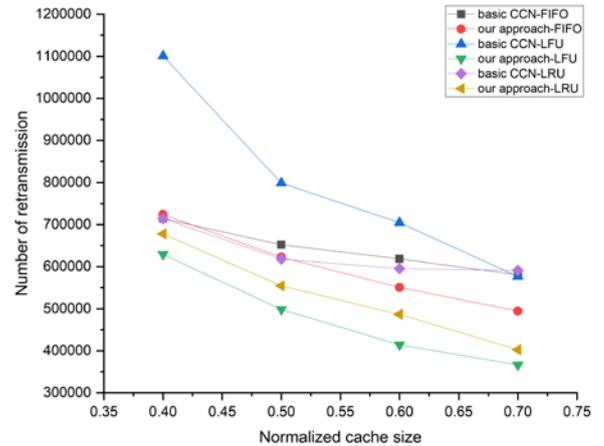

Fig. 13. The number of retransmissions for best route.

For best router, Fig. 14 shows the average response time for all routers in CCN to obtain content responses after forwarding interest packets. Our approach improves the accuracy of the routing information, so the waiting time caused by routing information failure is effectively reduced.

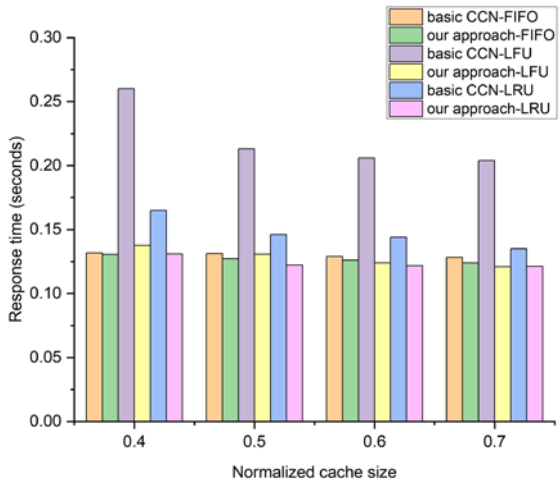

Fig. 14. Average response time for best route.

## VI. Conclusion

In CCN, cache miss causes numerous propagated packets to obtain the requested content that deteriorates network congestion and delay the response time of consumers. In this work, we have presented a query-based routing table update mechanism. The improved routing table allows interest packets to additionally query routers for the query name, and return the inquiry result through the data packet. The purpose is to real-time explore the cached content in routers. The update mechanism of the FIB takes into account the response time and existing time of routing information, so as to replace invalid routing information in time and ensure the accuracy of the routing information in the FIB. The basic CCN and our approach are established on the ndnSIM simulator. We have evaluated the performance of basic CCN and our optimization mechanism, for different normalized cache sizes from 0.4 to 0.7. The forwarding strategies adopted smart flooding and best route. Experiments are carried out for LRU, LFU, and FIFO. The experiment tested the number of flooding, the number of interest packets, the response time, and the number of retransmissions. The results show that our approach improves the response time of consumers and network congestion and is compatible with general forwarding strategies.

For future work, We will evaluate several other approaches that are used to determine which item in the PIT is selected as the query name. In addition, we will further study how to reduce the redundancy of the cached content between routers.